\documentclass[aps,prl,showpacs,twocolumn,amsmath,amssymb,superscriptaddress]{revtex4-1}
\usepackage{graphicx}
\usepackage{bm}
\usepackage{mathptmx}
\usepackage[english]{babel}
\usepackage{indentfirst}
\usepackage{color}
\usepackage{listings}
\usepackage{multibib}

\begin{document}

\title{Two-dimensional electron gas in the regime of strong light-matter coupling: Dynamical conductivity and all-optical measurements of Rashba and Dresselhaus coupling}

\author{Dmitry Yudin} 
\email{dimafizmath@ya.ru}
\affiliation{ITMO University, Saint Petersburg 197101, Russia}
\affiliation{Division of Physics and Applied Physics, Nanyang Technological University, Singapore 637371, Singapore}

\author{Ivan A. Shelykh}
\affiliation{ITMO University, Saint Petersburg 197101, Russia}
\affiliation{Division of Physics and Applied Physics, Nanyang Technological University, Singapore 637371, Singapore}
\affiliation{Science Institute, University of Iceland, IS-107 Reykjavik, Iceland}

\date{\today}

\begin{abstract}
A nonperturbative interaction of an electronic system with a laser field can substantially modify its physical properties. In particular, in two-dimensional (2D) materials with a lack of inversion symmetry, the achievement of a regime of strong light-matter coupling allows direct optical tuning of the strength of the Rashba spin-orbit interaction (SOI). Capitalizing on these results, we build a theory of the dynamical conductivity of a 2D electron gas with both Rashba and Dresselhaus SOIs coupled to an off-resonant high-frequency electromagnetic wave. We argue that strong light-matter coupling modifies qualitatively the dispersion of the electrons and can be used as a powerful tool to probe and manipulate the coupling strengths and adjust the frequency range where optical conductivity is essentially nonzero.
\end{abstract}

\pacs{72.40.+w,73.23.-b,71.70.Ej}
\maketitle

{\it Introduction.} Since the appearance of the pioneering works on spintronics, followed by unprecedented research progress in the field \cite{Spintronics}, there has been tremendous interest in studying spin-orbit coupled systems. This is mainly motivated by the possibility to use spin-orbit interactions (SOIs) for the design of prospective nanoelectronic devices \cite{Manchon2015} where the spin of a system can be manipulated without application of an external magnetic field. In two-dimensional (2D) electronic systems, SOIs can be provided either by a lack of inversion symmetry of the crystalline lattice itself (the so-called Dresselhaus term \cite{Dresselhaus1955}) or structural asymmetry of the quantum well (the Rashba term \cite{Bychkov1984}). While the strength of the Dresselhaus term is determined exclusively by the material and geometry of the structure, the strength of the Rashba term can be tuned by application of a gate voltage, which opens a way for the design of various spintronic components including Datta-Das spin field-effect transistors \cite{Datta1990}.

Meanwhile, the search for alternative ways to manipulate spin-orbit coupling still attracts considerable attention. It was recently proposed that the latter can be achieved by coupling of a 2D electron system with a strong off-resonant electromagnetic field (dressing field) \cite{Sheremet2016}, when no real absorption of the wave takes place but the spectrum of the system is changed. This corresponds to the so-called regime of strong light-matter coupling. The resulting dispersion renormalization was recently studied for the electrons in bulk semiconductors \cite{Vu2004,Vu2005}, quantum wells \cite{Mysyrovich1986,Wagner2010,Teich2013,Morina2015}, and graphene \cite{Glazov2014,Usaj2014,Oka2009,Oka2009a,Oka2011,Yudin2015,Kristinsson2016}. The dressing field also has a profound impact on the transport properties of low-dimensional electronic structures. In particular, it leads to an increase of dc conductivity of a two-dimensional electron gas (2DEG) and suppress the effect of weak localization \cite{Morina2015}. The oscillating behavior of conductivity and its strong anisotropy also have been predicted for monolayer graphene dressed by linearly polarized light \cite{Kristinsson2016} and three-dimensional topological insulators \cite{Shao2013}. Moreover, in graphene, a time-periodic circularly polarized field gives rise to a dynamical gap opening and resulting photocurrent can flow without any applied bias voltage \cite{Syzranov2008}.

In this Rapid Communication we investigate the effect of electromagnetic dressing on the transport properties of 2DEG with both Rashba and Dresselhaus SOIs in a quantum well grown in the [001] direction. Interestingly, the trade-off between Rashba and Dresselhaus couplings leads to a finite-frequency response with spectral features that are significantly different from those of a pure Rashba or Dresselhaus system \cite{Cruz2006}. The plasmon spectrum in Rashba-Dresselhaus systems also changes dramatically \cite{Badalyan2009,Cruz2014}. Thus, coupling to a strong off-resonant field opens possible ways of controlling the charge and spin current response of the system, providing also a tool to extract the Rashba and Dresselhaus couplings in all-optical measurements.

{\it Model.} We consider a spin-orbit coupled two-dimensional electron gas in which the electrons are restricted to move within a plane perpendicular to the $\hat{\mathbf{z}}$ axis irradiated by an external electromagnetic wave propagating perpendicular to the interface, $\mathbf{E}=\mathbf{E}_0\cos\Omega t$, where $E_0=\vert\mathbf{E}_0\vert$ is an amplitude of the wave and $\Omega$ is frequency. Periodic time dependence is characterized by a symmetry operation that corresponds to a translation by a period of a driving field, and Floquet quasienergies \cite{Grifoni1998,Platero2004,Kohler2005} describe the total phase shifts the quantum system picks up, evolving over a period. As long as the frequency of the irradiating field is far from the resonant frequencies of electronic interband transitions, so that interband absorption does not happen, and is high enough to satisfy a condition $\Omega\tau\gg 1$ (where $\tau$ stands for the relaxation time of a bare system), the problem can be mapped to an effective time-independent model in which the parameters of the undriven Hamiltonian are renormalized by the field. In our further discussion we focus on the linearly polarized dressing field only,  $\mathbf{E}_0=-E_0\hat{\mathbf{y}}$.

{\it Effective time-independent Hamiltonian.} We start our analysis with the Hamiltonian of a spin-orbit coupled two-dimensional electron system, 
\begin{equation}\label{inham}
H=\frac{p_x^2+p_y^2}{2m}+\alpha\left(p_y\sigma_x-p_x\sigma_y\right)+\beta\left(p_x\sigma_x-p_y\sigma_y\right),
\end{equation}

\noindent where $\sigma=\left(\sigma_x,\sigma_y,\sigma_z\right)$ is a vector of Pauli matrices acting in spin space, and constants $\alpha$ and $\beta$ characterize the strengths of Rashba and Dresselhaus couplings, respectively. This Hamiltonian describes, for example, an InAs-based quantum well grown in the [001] direction \cite{Giglberger2007}. The eigenstates of this Hamiltonian are purely determined by the electron momentum $\mathbf{p}=\left(p_x,p_y\right)$ and chirality of the spin branches. In the presence of an external electromagnetic field the Hamiltonian acquires a time-dependent term via a canonical replacement $\mathbf{p}\rightarrow\mathbf{p}-e\mathbf{A}(t)/c$, which originates from a minimal coupling to the field, where $\mathbf{A}(t)=-c\int^t\mathbf{E}(t^\prime)dt^\prime$ (here, $c$ is the speed of light). 

Performing unitary transformation with a matrix, 

\begin{gather}\nonumber
U(t)=\frac{1}{\sqrt{2}}e^{-i\frac{e^2E_0^2t}{4m\hbar\Omega^2}-i\frac{ep_yE_0}{m\hbar\Omega^2}\cos\Omega t+i\frac{e^2E_0^2}{8m\hbar\Omega^3}\sin2\Omega t} \\ \label{unitary}
\times\left(\begin{array}{cc}
e^{-i\gamma\cos\Omega t} & e^{i\gamma\cos\Omega t} \\
e^{-i\xi-i\gamma\cos\Omega t} & -e^{-i\xi+i\gamma\cos\Omega t}
\end{array}\right),
 \end{gather}

\noindent where $\gamma=eE_0\sqrt{\alpha^2+\beta^2}/(\hbar\Omega^2)$ is dimensionless field-matter coupling and $\tan\xi=\beta/\alpha$, and keeping zeroth-order harmonics \cite{supp} in the Floquet expansion only (which is possible for off-resonant external fields), we can reduce the problem to an effective time-independent Hamiltonian that resembles a bare Hamiltonian with effective anisotropic Rashba and Dresselhaus couplings renormalized by the field,

\begin{equation}\label{finham}
\tilde{H}=\frac{p_x^2+p_y^2}{2m}+\left(\alpha_yp_y+\beta_xp_x\right)\sigma_x-\left(\alpha_xp_x+\beta_yp_y\right)\sigma_y,
\end{equation}

\noindent where Rashba

\begin{equation}\label{anisr}
\alpha_x=\alpha\left[1-\frac{\alpha^2-\beta^2}{\alpha^2+\beta^2}\left(1-J_0(2\gamma)\right)\right], \quad \alpha_y=\alpha,
\end{equation}

\noindent and Dresselhaus-type couplings

\begin{equation}\label{anisd}
\beta_x=\beta\left[1+\frac{\alpha^2-\beta^2}{\alpha^2+\beta^2}\left(1-J_0(2\gamma)\right)\right], \quad \beta_y=\beta.
\end{equation}

\begin{figure}[htbp!]
\includegraphics[scale=0.315]{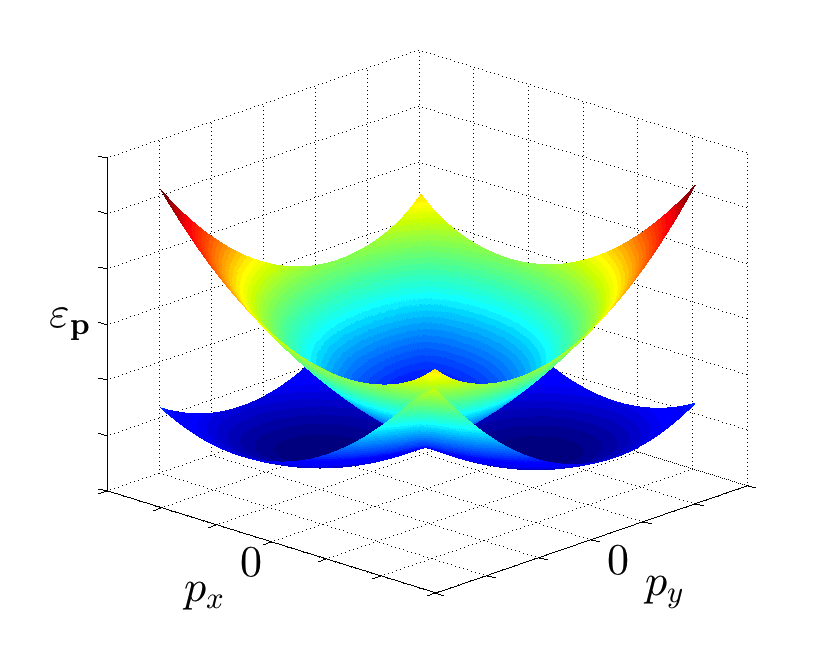}
\caption{The proposed renormalization of the dispersion relation $\varepsilon_{\mathbf{p}+}$ is illustrated schematically: The upper surface corresponds to 2DEG with both Rashba and Dresselhaus SOIs with no external field, and the lowest one results from renormalization by field with (\ref{anisr}) and (\ref{anisd}). To make the effect of renormalization more pronounced in (\ref{eigen}) we put $2m=1$, $\beta/\alpha=12/15$, and $\gamma=0.2$.}
\label{fig:band}
\end{figure}

\noindent In expressions (\ref{anisr}) and (\ref{anisd}) $J_0(2\gamma)$ is the zeroth-order Bessel function of the first kind. Thus, the off-resonant external electromagnetic field provides a versatile tool to tune the corresponding spin-orbit strengths. One can easily verify that the dispersion relations of dressed electrons and corresponding eigenstates (see Fig.~\ref{fig:band} to observe renormalization due to an external field) of the Hamiltonian $\tilde{H}$ are determined by

\begin{equation}\label{eigen}
\varepsilon_{\mathbf{p}\lambda}=\frac{p^2}{2m}+\lambda p\Delta(\theta), \quad \vert\mathbf{p}\lambda\rangle=\frac{1}{\sqrt{2}}\left(\begin{array}{c}
1 \\ \lambda e^{-i\Phi}
\end{array}\right),
\end{equation}

\noindent where $\tan\Phi=\left(\alpha_x\cos\theta+\beta_y\sin\theta\right)/\left(\alpha_y\sin\theta+\beta_x\cos\theta\right)$, $\lambda$, the chirality index is denoted by $\lambda=\pm1$, and the anisotropic spin splitting is defined by

\begin{equation}\label{split}
\Delta(\theta)=\sqrt{\left(\alpha_x\cos\theta+\beta_y\sin\theta\right)^2+\left(\alpha_y\sin\theta+\beta_x\cos\theta\right)^2}.
\end{equation}

\noindent It is worth noting that the angle $\tan\theta=p_y/p_x$, while for a system doped up to $E_F>0$ and a concentration of charge carriers $n$,

\begin{equation}\label{fermi}
E_F=\frac{\pi n\hbar^2}{m}-\frac{m\left(\alpha_x^2+\alpha_y^2+\beta_x^2+\beta_y^2\right)}{2},
\end{equation}

\noindent can be tuned by changing the field parameters. Note that the expression for $\tilde{H}$ and consequent relations are formally true as long as the argument of $J_0(2\gamma)$ is far from the nulls of the Bessel function \cite{supp}.

{\it Optical conductivity.} The effects of electromagnetic dressing can be experimentally explored by studying the optical response of the system in a pump-probe geometry. In this case, a sample is excited by a continuous-wave highly intense laser (pump) while the second pulse (probe) is used for characterization of the excited states of the hybrid light-matter system. The proper description of experimentally relevant signals requires an adequate understanding of how an electromagnetic pulse of finite length propagates through a material which is driven out of equilibrium. If the probing field is weak enough, its effect shows up as a linear response of the current to the external field $\delta j_a(\omega)=\sigma_{ab}(\omega)\delta E_b(\omega)$, where the subscripts stand for Cartesian components of the vectors and tensors. It is worth noting that only the probe field is assumed to be weak, whereas no assumptions have been made about the strength of the pump field. Along with a standard Drude peak, the conductivity of a spin-orbit coupled system picks up an extra term determined by the Kubo formula. For a probing field of frequency $\omega$ it can be evaluated as follows,

\begin{equation}\label{kubo}
\sigma_{ab}(\omega)=\frac{1}{\hbar\omega}\int\limits_0^\infty dt\langle\left[\hat{j}_a(t),\hat{j}_b(0)\right]\rangle e^{i(\omega+i\delta)t},
\end{equation}

\noindent where $\delta$ is a positive infinitesimal constant introduced to guarantee the convergence of the integral. The angular brackets stand for quantum and thermal averaging. 

With the help of the Hamiltonian $\tilde{H}$ we can estimate the current operators,

\begin{equation}\label{current}
\hat{\mathbf{j}}=\nabla_\mathbf{p}\tilde{H}=-\frac{e}{m}\left\lbrace\begin{array}{c}
p_x \\ p_y
\end{array}\right\rbrace-e\sigma_x\left\lbrace\begin{array}{c}
\beta_x \\ \alpha_y
\end{array}\right\rbrace
+e\sigma_y\left\lbrace\begin{array}{c}
\alpha_x \\ \beta_y
\end{array}\right\rbrace.
\end{equation}

\noindent Without loss of generality, in the following we assume $\omega>0$, and after quite straightforward algebra we obtain \cite{supp}:

\begin{gather}\nonumber
\mathrm{Re}\sigma_{ab}(\omega)=\frac{e^2\left(\alpha_x\alpha_y-\beta_x\beta_y\right)^2}{4\pi\omega\hbar^2}\int\frac{d^2p}{\Delta^2(\theta)}\times \\ \label{conduct}
\times\left(\begin{array}{cc}
\sin^2\theta & -\sin\theta\cos\theta \\
-\sin\theta\cos\theta & \cos^2\theta
\end{array}\right)\delta\left(\varepsilon_{\mathbf{p}+}-\varepsilon_{\mathbf{p}-}-\hbar\omega\right).
\end{gather}

\begin{figure}[htbp!]
	\begin{minipage}[t]{0.55\textwidth}
	\includegraphics[width=\textwidth]{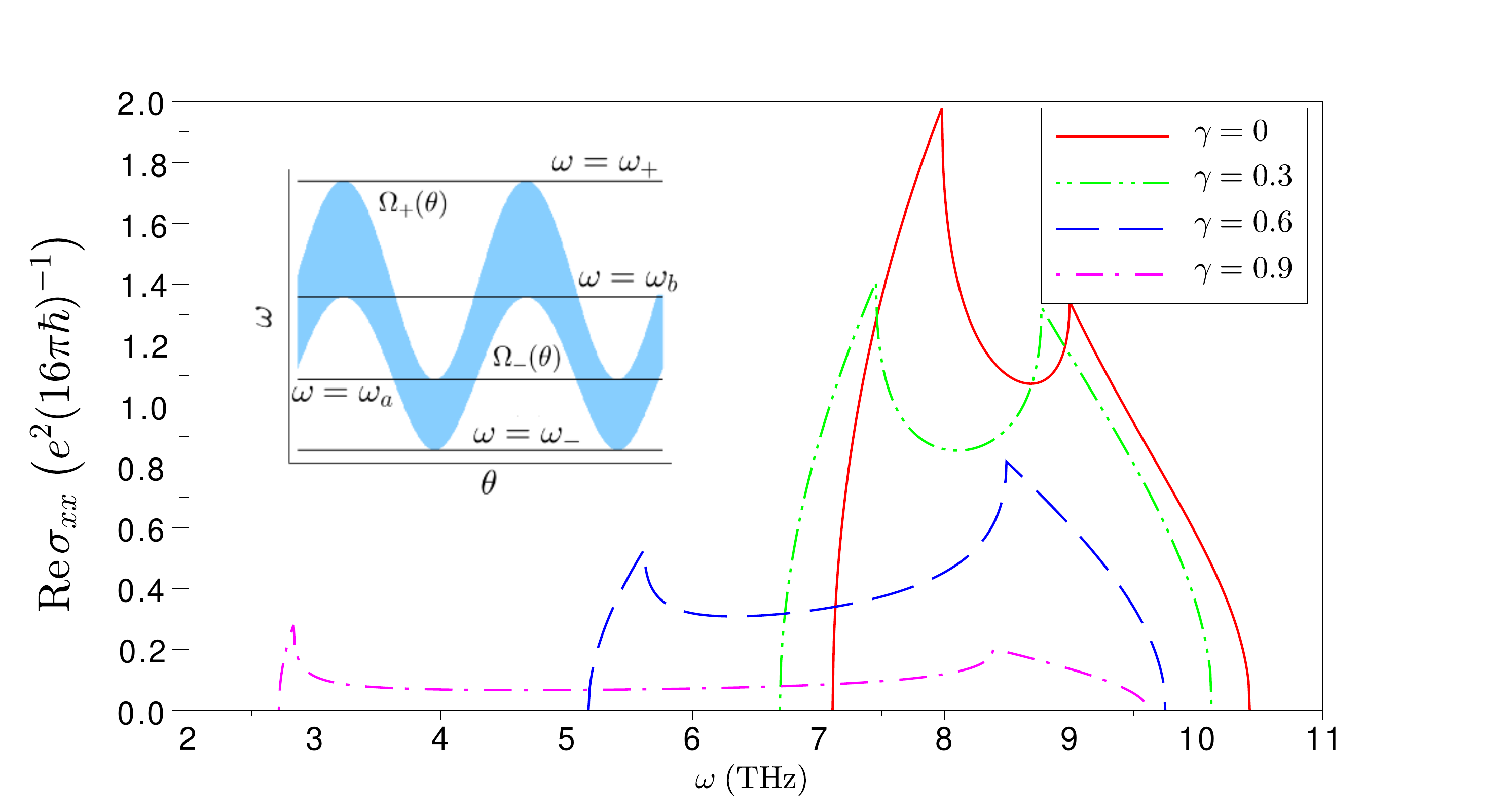}
	\end{minipage}
	\begin{minipage}[t]{0.55\textwidth}
	\includegraphics[width=\textwidth]{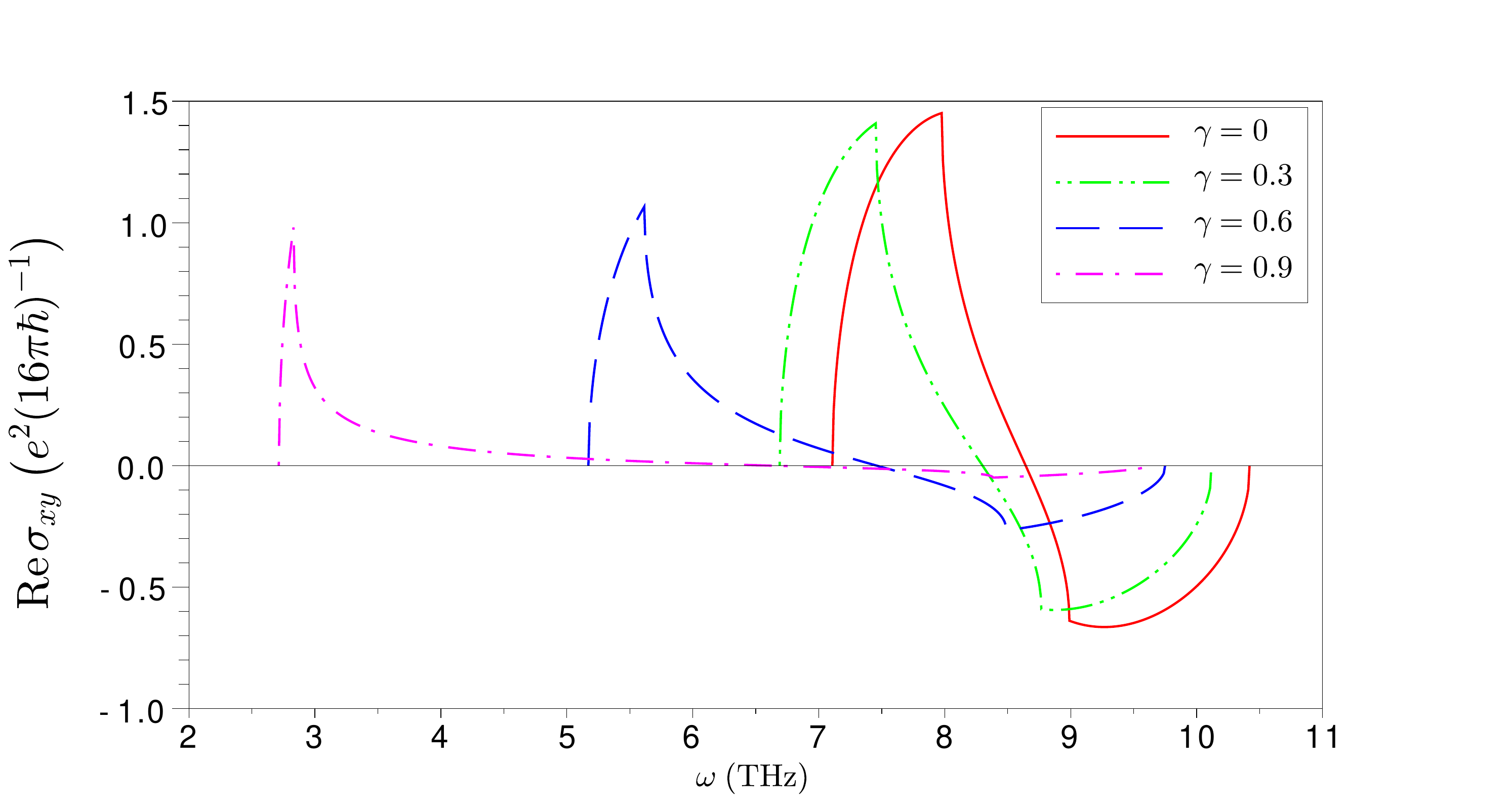}
	\end{minipage}
\caption{Components of the conductivity tensor (\ref{conduct}) $\mathrm{Re}\sigma_{xx}(\omega)$ and $\mathrm{Re}\sigma_{xy}(\omega)$ plotted for different values of the field-matter coupling $\gamma=eE_0\sqrt{\alpha^2+\beta^2}/(\hbar\Omega^2)$ with a fixed ratio $\beta/\alpha=0.125$. The red solid line specifies the case with no dressing. An increase in the field-matter coupling results in the frequency domain being broadened and shifting slightly past the original position. The dashed area in the inset to the top panel shows the integration area in (\ref{conduct}), while the two peaks on the main plots correspond precisely to the frequencies $\omega=\omega_a$ and $\omega=\omega_b$.}
\label{fig:conductivity}
\end{figure}

\noindent Expression (\ref{conduct}) clearly manifests that the conductivity due to spin-orbit coupling disappears for $\vert\alpha\vert=\vert\beta\vert$. In fact, in this case a delicate interplay between the Dresselhaus and the Rashba couplings leads to a momentum-independent eigenspinor  \cite{Schliemann2003a,Schliemann2003b}, and the conductivity of the system becomes isotropic. Analytical formulas for $\mathrm{Re}\sigma_{ab}(\omega)$ are listed in Ref. \cite{supp}. In contrast to a pure Rashba or Dresselhaus system, the frequency range where $\mathrm{Re}\sigma_{ab}(\omega)\neq0$ is more broadened $\omega_-\leq\omega\leq\omega_+$ (Fig.~\ref{fig:conductivity}), where

\begin{equation}\label{frequen}
\hbar\omega_\pm=\hbar\Omega_\pm(\Delta_\pm),
\end{equation}

\noindent and we have defined the functions

\begin{equation}\label{freq}
\hbar\Omega_\pm(\Delta(\theta))=2\Delta(\theta)\Big(\sqrt{m^2\Delta^2(\theta)+2mE_F}\pm m\Delta(\theta)\Big),
\end{equation}

\noindent which determine the integration area $\Omega_-(\theta)\leq\omega\leq\Omega_+(\theta)$ over the polar angle $\theta$, while $\Delta_\pm$ denote the maximum and minimum of $\Delta(\theta)$, respectively.  The energies $\hbar\omega_\pm$ correspond to the minimum and the maximum photon energy required to induce the optical transitions between the initial $\lambda=-1$ and final $\lambda=+1$ subbands and coincide with the absorption edges of the spectrum.

{\it Results and discussion.} Close inspection of formulas (\ref{frequen}) reveals that the presence of an intense electromagnetic field allows one to optically tune the values of $\hbar\omega_\pm$. The results of the numerical calculations of conductivity (\ref{conduct}) are shown in Fig.~\ref{fig:conductivity}. We used the parameters that are experimentally accessible in InAs-based quantum wells grown in the [001] direction \cite{Giglberger2007,Bastidas2007,Ganichev2014}, $m=0.055m_e$, where $m_e$ is the free-electron mass, $\alpha=1.6\times 10^{-9}$ eV cm, $\beta=0.125\alpha$, $n=5\times 10^{11}$ cm$^{-2}$, and $\gamma=$0, 0.3, 0.6, and 0.9.  It can be seen that $\mathrm{Re}\sigma_{ab}(\omega)$ is non-zero only in a well defined interval of frequencies, $\omega_-<\omega<\omega_+$, and both $\omega_+$ and $\omega_-$ decrease as functions of dimensionless field-matter coupling $\gamma$, and the range of the frequencies where the real part of the conductivity is nonzero becomes broadened. The major factor that determines this interval is related to the Fermi surface topology and explains the possible excitation energies of electron-hole pairs (electron-hole continuum). Note that for $\gamma=0$ our results coincide with those reported in previous studies \cite{Cruz2006}.

\begin{figure}[htbp!]
	\begin{minipage}[t]{0.54\textwidth}
	\includegraphics[width=\textwidth]{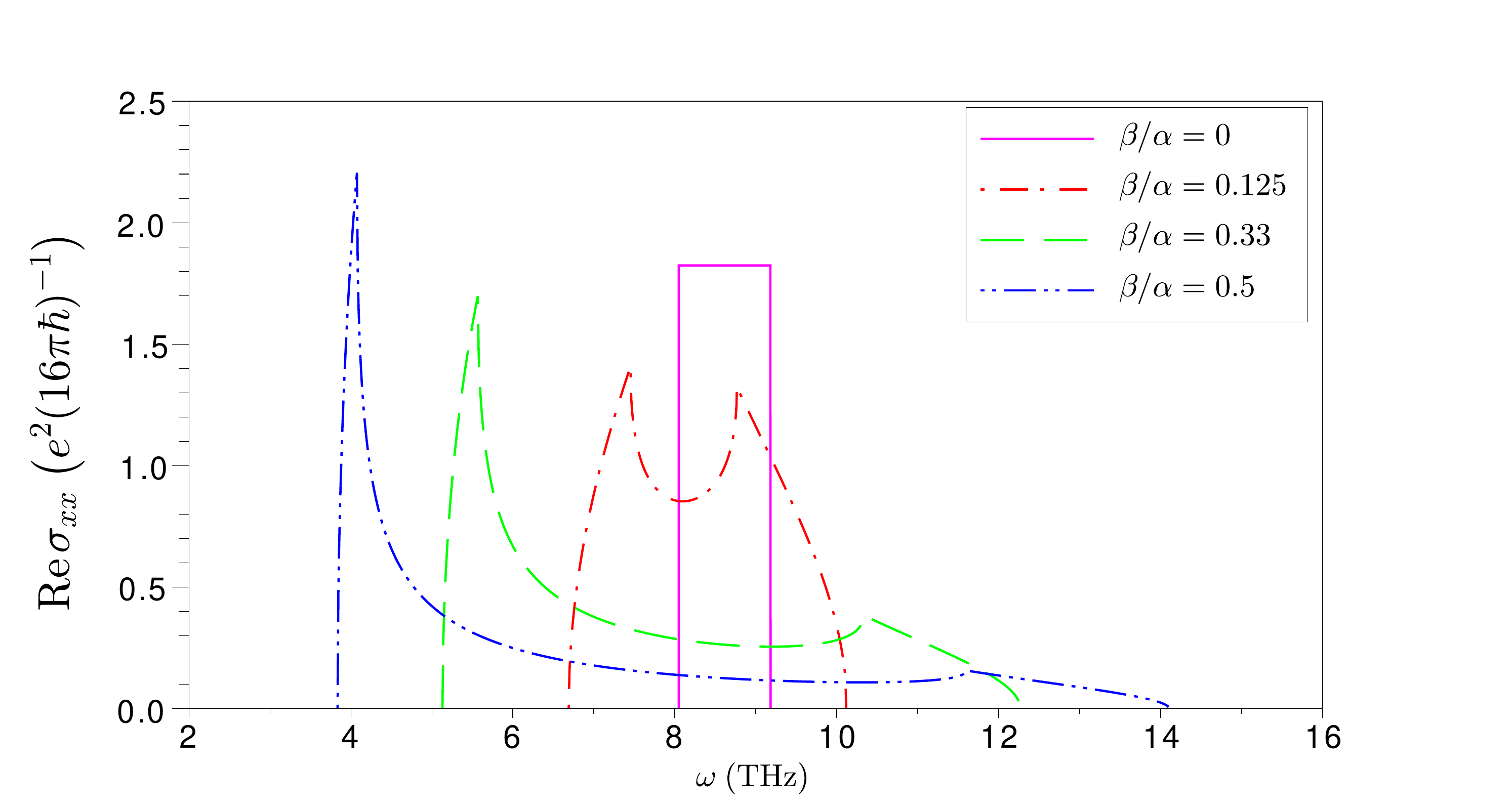}
	\end{minipage}
	\begin{minipage}[t]{0.54\textwidth}
	\includegraphics[width=\textwidth]{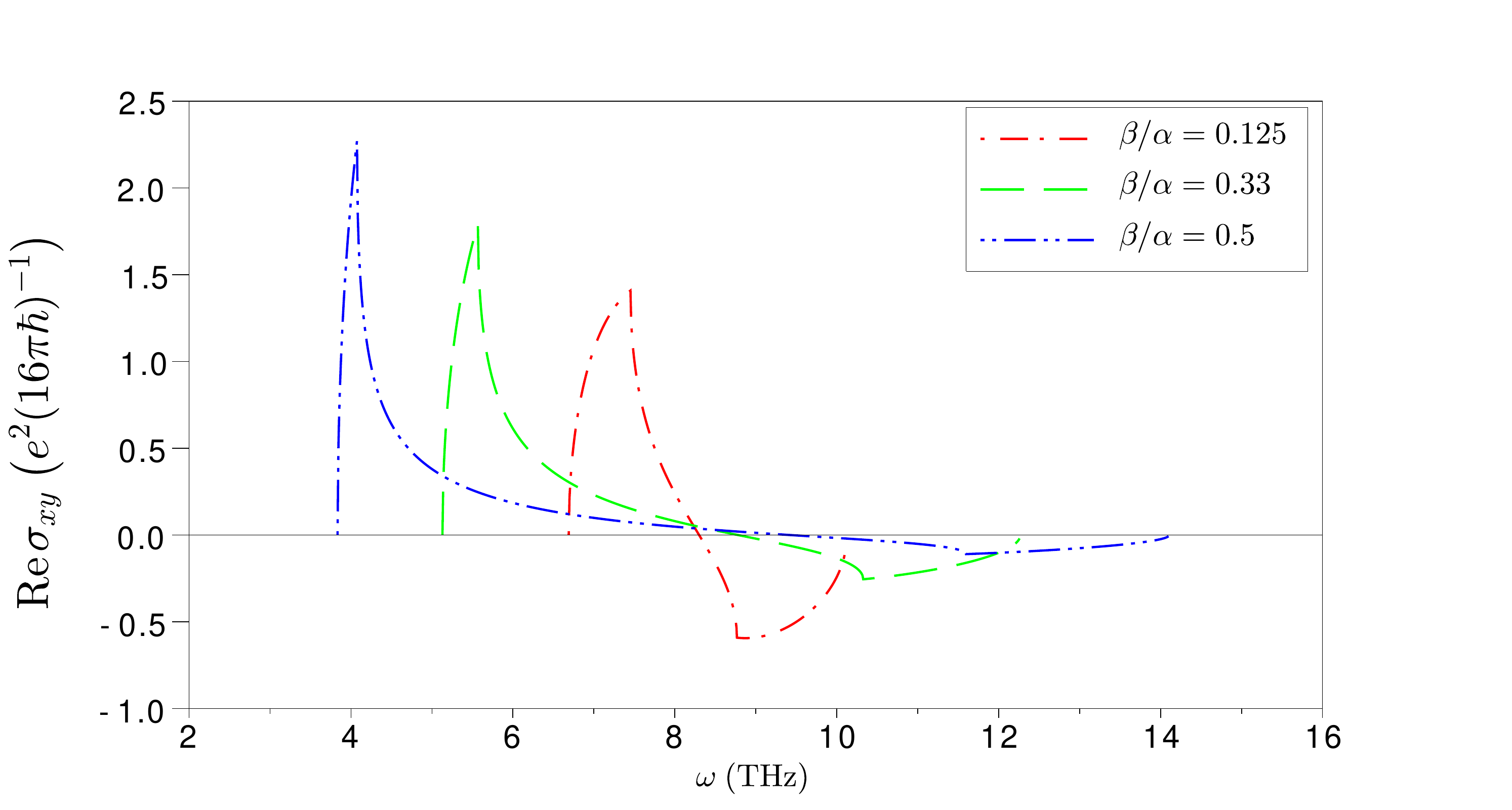}	
	\end{minipage}
\caption{Components of the conductivity tensor (\ref{conduct}) $\mathrm{Re}\sigma_{xx}(\omega)$ and $\mathrm{Re}\sigma_{xy}(\omega)$ plotted for the fixed field-matter coupling $\gamma=0.3$ and the relative strength of spin-orbit coupling $\beta/\alpha=$ 0.125, 0.33, and 0.5. The two-peak structure outlined in the main text is clearly visible, allowing for independent experimental determinations of the Rashba and Dresselhaus couplings. The conductivity of a pure Rashba system is plotted in the top panel (solid line). One can observe that a pure Rashba (or Dresselhaus) structure is characterized by a constant value in a more narrow interval of frequency domain.}
\label{fig:conductivity1}
\end{figure}

The two peaks in Figs.~\ref{fig:conductivity} and \ref{fig:conductivity1} correspond to electronic excitations involving states with allowed wave vectors exactly at $\hbar\omega_a=\hbar\Omega_+(\Delta_-)$ and $\hbar\omega_b=\hbar\Omega_-(\Delta_+)$ (featured in the inset to the Fig.~\ref{fig:conductivity}), provided $\omega_-<\omega_a<\omega_b<\omega_+$. This is in huge contrast to the results of a pure Rashba or Dresselhaus system for which $\mathrm{Re}\sigma_{xx}(\omega)=e^2\vert J_0(2\gamma)\vert/(16\pi\hbar)$, and $\mathrm{Re}\sigma_{yy}(\omega)=e^2/(16\pi\hbar\vert J_0(2\gamma)\vert)$, in the finite frequency range determined by $\vert\hbar\omega-2\alpha\sqrt{m^2\alpha^2+2mE_F}\vert\leq 2m\alpha^2$.

One can also observe that the lower peak of the components of the conductivity tensor $\mathrm{Re}\sigma_{ab}(\omega)$ at $\omega=\omega_a$ moves towards $\omega_-$ with an increase of $\gamma$ (see Fig.~\ref{fig:conductivity}). This effect becomes even more pronounced when the ratio $\beta/\alpha$ grows. Interestingly, $\mathrm{Re}\sigma_{xx}(\omega)$ reaches maximal value in the absence of a dressing field and becomes suppressed with an increase of $\gamma$. This is in contrast to the behavior of $\mathrm{Re}\sigma_{yy}(\omega)$, which is shown to take the lowest value in the absence of the field and gains a maximum value at $\omega_-$ with increasing $\gamma$ (not shown).  

Contrary to a pure Rashba or Dresselhaus material, in which it requires a circularly polarized field \cite{Ojanen2012}, in a biased two-dimensional electron gas the presence of both couplings leads to the emergence of Hall-type conductivity of the charge carriers, even in the absence of an external magnetic field \cite{Bryksin2006} (see also Ref. \cite{supp}). Results presented in Fig. \ref{fig:conductivity} show that the Hall-type conductivity is also quite sensitive to the dressing field. The off-diagonal components of the frequency-dependent conductivity tensor can be accessed, e.g., via measurements of the Faraday rotation angle, which for sufficiently thin films is proportional to $\sigma_{xy}(\omega)$ (see, e.g., Ref. \cite{Volkov1985}).

It should be noted that the modification of the dynamical conductivity by a dressing field allows for an experimental determination of the relative strength of the spin-orbit coupling $\beta/\alpha$. The current methods include photocurrent measurements \cite{Ganichev2004,Giglberger2007} or optical monitoring of electron spin precession \cite{Meier2007}, or persistent charge and spin current measurements in a mesoscopic ring \cite{Maiti2011}. We propose to extract $\alpha$ and $\beta$ from spectroscopic experiments in the pump-probe regime. Based on the theory developed in this Rapid Communication, one can show that 

\begin{equation}
E_F=\left(p_0^2-\sqrt{p_0^4-Am^2}\right)/(4m),
\end{equation}

\begin{equation}
\vert\alpha^2-\beta^2\vert=\frac{B}{8mE_F\vert J_0(2\gamma)\vert},
\end{equation}

\noindent and

\begin{equation}
\alpha^2+\beta^2=\frac{A\vert J_0(2\gamma)\vert+2\sqrt{A^2J_0^2(2\gamma)+8B^2\left(1-J_0^2(2\gamma)\right)}}{32mE_F\vert J_0(2\gamma)\vert},
\end{equation}

\noindent where $p_0=\sqrt{2\pi n\hbar^2}$ is the Fermi momentum of a spin-degenerate two-dimensional electron gas, $A=\hbar^2\left(\omega_-\omega_a+\omega_b\omega_+\right)$, and $B=\hbar^2\sqrt{\omega_-\omega_a\omega_b\omega_+}$. The parameters $\omega_\pm$, $\omega_a$, and $\omega_b$ can be extracted implicitly from the experimentally measured dynamical conductivity curves for various light-matter coupling parameters $\gamma$ (Fig.~\ref{fig:conductivity1}).

{\it Conclusions and outlook.} In this Rapid Communication we have provided a systematic and self-contained analysis of the transport properties of a dressed 2D electron system with simultaneous Rashba and Dresselhaus SOIs. We showed that strong light-matter coupling leads to renormalization of the spectrum of the system, which results in a dramatic modification of the dynamical conductivity of a system. In particular, we demonstrated that the frequency range where the conductivity is essentially nonzero can be tuned by properly adjusting the parameters of the dressing field. Moreover, we proposed a way to define independently the constants of Rashba and Dresselhaus SOIs in all-optical measurements. 
 
We acknowledge support from the Singaporean Ministry of Education under AcRF Tier 2 Grant No. MOE2015-T2-1-055 and Ministry of Education and Science of the Russian Federation under Increase Competitiveness Program 5-100. I.A.S. thanks Horizon2020 ITN NOTEDEV and RANNIS excellence Grant No. 163082-051.

\onecolumngrid
\setcounter{page}{0}
\setcounter{table}{0}
\setcounter{section}{0}
\setcounter{figure}{0}
\setcounter{equation}{0}
\renewcommand{\thepage}{\Roman{page}}
\renewcommand{\thesection}{S\arabic{section}}
\renewcommand{\thetable}{S\arabic{table}}
\renewcommand{\thefigure}{S\arabic{figure}}
\renewcommand{\theequation}{S\arabic{equation}}
\cleardoublepage
\vfill\eject
%\thispagestyle{empty}
%\phantom{hi}\vfill\eject
\section{Supplemental Material}
\section{Derivation of the effective Hamiltonian}

In this section we provide a basic sketch of the effective time-independent Hamiltonian computation. In the presence of an external electromagnetic field $\mathbf{E}=\mathbf{E}_0\cos\Omega t$, that in the following is assumed to be linearly polarized $\mathbf{E}_0=-E_0\hat{\mathbf{y}}$, of the amplitude $E_0$ and the frequency $\Omega$ the Hamiltonian of a two-dimensional electron gas with account of both Rashba and Dresselhaus spin-orbit couplings is expressed as the sum

\begin{equation}\label{s1}
H(t)=\frac{1}{2m}\left(\mathbf{p}-\frac{e}{c}\mathbf{A}(t)\right)^2+\left(\alpha\sigma_x-\beta\sigma_y\right)\left(p_y-\frac{e}{c}A_y(t)\right)-\left(\alpha\sigma_y-\beta\sigma_x\right)p_x.
\end{equation}

\noindent We have chosen the Cartesian system with the $\hat{\mathbf{z}}-$axis perpendicular to the plane of electron motion. The formula (\ref{s1}) is purely determined by the corresponding electron momentum $\mathbf{p}=\left(p_x,p_y\right)$, the spin-orbit couplings strengths $\alpha$ (Rashba) and $\beta$ (Dresselhaus), as well as a set of Pauli matrices $\bm{\sigma}=\left(\sigma_x,\sigma_y,\sigma_z\right)$ acting in spin space. The time-dependent contribution originates from a minimal coupling to the field, we use a spatially uniform gauge for the vector potential $\mathbf{A}(t)=-c\int^t\mathbf{E}(t^\prime)dt^\prime$ in the following (where $c$ is the speed of light). For the given field polarization the Hamiltonian fully describes evolution $i\hbar\partial_t\Psi(t)=H(t)\Psi(t)$ of the system. Having been applied unitary transformation

\begin{equation}
U(t)=\frac{1}{\sqrt{2}}\left(\begin{array}{cc}
e^{-i\gamma\cos\Omega t} & e^{i\gamma\cos\Omega t} \\
e^{-i\xi-i\gamma\cos\Omega t} & -e^{-i\xi+i\gamma\cos\Omega t}
\end{array}\right)e^{-i\frac{e^2E_0^2t}{4m\hbar\Omega^2}-i\frac{ep_yE_0}{m\hbar\Omega^2}\cos\Omega t+i\frac{e^2E_0^2}{8m\hbar\Omega^3}\sin2\Omega t},
\end{equation}

\noindent to the Hamiltonian (\ref{s1}), where $\gamma=\frac{eE_0\sqrt{\alpha^2+\beta^2}}{\hbar\Omega^2}$, the effective dimensionless field-matter coupling, and $\tan\xi=\frac{\beta}{\alpha}$ determines the relative strength of spin-orbit interaction, we finally end up with dynamic equations

\begin{equation}\label{td2}
i\hbar\dfrac{d\Phi_\pm}{dt}=\left(\frac{p^2}{2m}\pm\sqrt{\alpha^2+\beta^2}\left(p_y+p_x\sin2\xi\right)\right)\Phi_\pm\mp i\Phi_\mp p_x\cos2\xi\sqrt{\alpha^2+\beta^2}e^{\pm 2i\gamma\cos\Omega t},
\end{equation}

\noindent for the components $\Phi(t)=\left(\Phi_+(t),\;\Phi_-(t)\right)^T=U^\dagger(t)\Psi(t)$. A set of equations (\ref{td2}) describes evolution of charge carriers in a two-dimensional Rashba-Dresselhaus electron gas irradiated with the external linearly polarized light propagating normally to its plane. Remarkably, exponentials of trigonometric function on the right-hand side of the equations (\ref{td2}) are simplified with the aid of Jacobi-Anger expansion: $e^{iz\cos\eta}=\sum\limits_{n=-\infty}^\infty i^nJ_n(z)e^{in\eta}$, where $J_n(z)$ is the $n-$th order Bessel function of the first kind, and the Floquet Hamiltonian in the extended Hilbert space is defined by

\begin{equation}
\left(\begin{array}{ccccc}
\ddots & & & & \ddots \\
 & H_{00} & \ldots & H_{0n} & \\
 & \vdots & \ddots & \vdots & \\
 & H_{n0} & \ldots & H_{nn} & \\
\ddots & & & & \ddots
\end{array}\right),
\end{equation}

\noindent where

\begin{equation}
H_{ln}=\left(\begin{array}{cc}
\left(\frac{p^2}{2m}+l\hbar\Omega+\sqrt{\alpha^2+\beta^2}\left(p_y+p_x\sin2\xi\right)\right)\delta_{ln} & -i^{l-n+1}p_x\cos2\xi J_{l-n}(2\gamma)\sqrt{\alpha^2+\beta^2} \\
i^{n-l+1}p_x\cos2\xi J_{l-n}(2\gamma)\sqrt{\alpha^2+\beta^2} & \left(\frac{p^2}{2m}+l\hbar\Omega-\sqrt{\alpha^2+\beta^2}\left(p_y+p_x\sin2\xi\right)\right)\delta_{ln}
\end{array}\right),
\end{equation}

\noindent here $\delta_{ln}$ is the Kronecker symbol. If treat off-diagonal elements as a small perturbation, i.e. we assume

\begin{equation}
\left\vert\frac{\left(\alpha^2-\beta^2\right)p_xJ_n^2(2\gamma)}{J_0(2\gamma)\left[\sqrt{\alpha^2+\beta^2}\left(n\hbar\Omega+2p_y\sqrt{\alpha^2+\beta^2}\right)+4\alpha\beta p_x\right]}\right\vert\ll1,
\end{equation}

\noindent which means that $J_0(2\gamma)$ has to be far from the nulls of the Bessel function and the frequency of a driving field $\Omega$ should be much larger than the absorption bandwidth, we can neglect higher order harmonics and keep $J_0(z)$ in Jacobi-Anger expansion exclusively. Thus, unless otherwise stated dynamics of quasiparticles dressed by the external field is governed by the effective Hamiltonian:

\begin{equation}\label{effham}
H_\mathrm{eff}=\left(\begin{array}{cc}
\frac{p^2}{2m}+\left(p_y+p_x\sin2\xi\right)\sqrt{\alpha^2+\beta^2} & -ip_x\sqrt{\alpha^2+\beta^2}J_0(2\gamma)\cos2\xi \\ \\
ip_x\sqrt{\alpha^2+\beta^2}J_0(2\gamma)\cos2\xi & \frac{p^2}{2m}-\left(p_y+p_x\sin2\xi\right)\sqrt{\alpha^2+\beta^2}
\end{array}\right),
\end{equation}

\noindent The Hamiltonian $H_\mathrm{eff}$ in the original basis ($\sigma_z$ is diagonal) can be written in a similar fashion as the initial Hamiltonian (\ref{s1}), namely

\begin{equation}\label{s5}
\tilde{H}=\frac{p^2}{2m}+\left(\alpha_yp_y+\beta_xp_x\right)\sigma_x-\left(\alpha_xp_x+\beta_yp_y\right)\sigma_y,
\end{equation}

\noindent with effective anisotropic Rashba

\begin{equation}
\bm{\alpha}=\left(\alpha_x,\,\alpha_y\right)=\left(\frac{\alpha}{\alpha^2+\beta^2}\Big(2\beta^2+\left(\alpha^2-\beta^2\right)J_0(2\gamma)\Big),\,\alpha\right),
\end{equation}

\noindent and Dresselhaus couplings

\begin{equation}
\bm{\beta}=\left(\beta_x,\,\beta_y\right)=\left(\frac{\beta}{\alpha^2+\beta^2}\Big(2\alpha^2-\left(\alpha^2-\beta^2\right)J_0(2\gamma)\Big),\,\beta\right).
\end{equation}

\noindent The Hamiltonian (\ref{s5}) characterizes the behavior of dressed electrons, which can be thought of as a composite particle of an electron and the electromagnetic field. It can be shown that the dispersion relation and corresponding eigenstates of the Hamiltonian $\tilde{H}$ are determined by

\begin{equation}\label{dispersion}
\varepsilon_{\mathbf{p}\lambda}=\frac{p^2}{2m}+\lambda p\Delta(\theta), \quad \vert\mathbf{p}\lambda\rangle=\frac{1}{\sqrt{2}}\left(\begin{array}{c}
1 \\ \lambda e^{-i\Phi}
\end{array}\right), \quad \tan\Phi=\frac{\alpha_x\cos\theta+\beta_y\sin\theta}{\alpha_y\sin\theta+\beta_x\cos\theta},
\end{equation}

\noindent

\noindent where $\lambda$ denotes chirality index $\lambda=\pm 1$, while anisotropic spin splitting energy is defined by $\Delta(\theta)=\sqrt{\left(\alpha_x\cos\theta+\beta_y\sin\theta\right)^2+\left(\alpha_y\sin\theta+\beta_x\cos\theta\right)^2}$.

\section{Optical conductivity: Computational details}

For a probing field of frequency $\omega$ we can evaluate the dynamical conductivity according to the Kubo formula:

\begin{equation}\label{cond1}
\sigma_{ab}(\omega)=\frac{1}{\hbar\omega}\int\limits_0^\infty dt\langle\left[\hat{j}_a(t),\hat{j}_b(0)\right]\rangle e^{i(\omega+i\delta)t},
\end{equation}

\noindent where $\delta$ is a positive infinitesimal constant introduced to guarantee the convergence of the right hand side. The angular brackets stand for quantum and thermal averaging

\begin{equation}\label{cond2}
\langle\left[\hat{j}_a(t),\hat{j}_b(0)\right]\rangle=\sum\limits_{\lambda\lambda^\prime}\int\frac{d^2p}{(2\pi\hbar)^2}\Big(f(\varepsilon_{\mathbf{p}\lambda})-f(\varepsilon_{\mathbf{p}\lambda^\prime})\Big)\langle\mathbf{p}\lambda\vert j_a\vert\mathbf{p}\lambda^\prime\rangle\langle\mathbf{p}\lambda^\prime\vert j_b\vert\mathbf{p}\lambda\rangle,
\end{equation}

\noindent here $f(x)=\left(e^{\beta x}+1\right)^{-1}$ denotes the Fermi-Dirac distribution function, while $\varepsilon_{\mathbf{p}\lambda}$ and $\vert\mathbf{p}\lambda\rangle$ are determined by (\ref{dispersion}). Thus,

\begin{equation}\label{cond3}
\sigma_{ab}(\omega)=\frac{i}{\omega}\sum\limits_{\lambda\neq\lambda^\prime}\int\frac{d^2p}{(2\pi\hbar)^2}\frac{f(\varepsilon_{\mathbf{p}\lambda})-f(\varepsilon_{\mathbf{p}\lambda^\prime})}{\hbar\omega+\varepsilon_{\mathbf{p}\lambda}-\varepsilon_{\mathbf{p}\lambda^\prime}+i\delta}\langle\mathbf{p}\lambda\vert j_a\vert\mathbf{p}\lambda^\prime\rangle\langle\mathbf{p}\lambda^\prime\vert j_b\vert\mathbf{p}\lambda\rangle.
\end{equation}

\noindent With the help of the Hamiltonian $\tilde{H}$ we can evaluate current operators

\begin{equation}\label{cur}
j_x=-e\frac{\partial H}{\partial p_x}=-e\left(\frac{p_x}{m}+\beta_x\sigma_x-\alpha_x\sigma_y\right), \quad
j_y=-e\frac{\partial H}{\partial p_y}=-e\left(\frac{p_y}{m}+\alpha_y\sigma_x-\beta_y\sigma_y\right).
\end{equation}

\noindent Without loss of generality in the following we assume $\omega>0$ and after quite straightforward algebra derive (in the clean limit, i.e., no impurity scattering is allowed)

\begin{equation}\label{cond4}
\mathrm{Re}\sigma_{ab}(\omega)=\frac{e^2\left(\alpha_x\alpha_y-\beta_x\beta_y\right)^2}{4\pi\omega\hbar^2}\int\frac{d^2p}{\Delta^2(\theta)}\left(\begin{array}{cc}
\sin^2\theta & -\sin\theta\cos\theta \\
-\sin\theta\cos\theta & \cos^2\theta
\end{array}\right)\delta\left(\varepsilon_{\mathbf{p}+}-\varepsilon_{\mathbf{p}-}-\hbar\omega\right),
\end{equation}

\noindent where the integration area is restricted by $\vert p-\sqrt{m^2\Delta^2(\theta)+2mE_F}\vert\leq m\Delta(\theta)$ (for the sake of simplicity, we work in the limit of vanishing temperature $T=0$). Due to delta-function in the integrand (\ref{cond4}) the integration over $\vert\mathbf{p}\vert$ leads to the replacement $\vert\mathbf{p}\vert=\frac{\hbar\omega}{2\Delta(\theta)}$, and we eventually derive:

\begin{equation}\label{cond5}
\mathrm{Re}\sigma_{ab}(\omega)=\frac{e^2\left(\alpha_x\alpha_y-\beta_x\beta_y\right)^2}{16\pi\hbar}\int\frac{d\theta}{\Delta^4(\theta)}\left(\begin{array}{cc}
\sin^2\theta & -\sin\theta\cos\theta \\
-\sin\theta\cos\theta & \cos^2\theta
\end{array}\right)\theta\left(\hbar\Omega_+(\theta)-\hbar\omega\right)\theta\left(\hbar\omega-\hbar\Omega_-(\theta)\right),
\end{equation}

\noindent where $\theta(x)$ is the stepwise function, and

\begin{equation}\label{b1}
\hbar\Omega_\pm(\theta)=2\Delta(\theta)\left[\sqrt{m^2\Delta^2(\theta)+2mE_F}\pm m\Delta(\theta)\right].
\end{equation}

\noindent If we define $f=\alpha_x^2+\beta_x^2$, $g=\alpha_x\beta_y+\alpha_y\beta_x$, and $h=\alpha_y^2+\beta_y^2$ the maximum $\Delta_+$ and the minimum $\Delta_-$ values of $\Delta(\theta)$ are specified by

\begin{equation}\label{deltaex}
\Delta_\pm^2=\frac{f+h}{2}\pm\frac{\sqrt{\left(f-h\right)^2+4g^2}}{2}.
\end{equation}

\noindent Therefore, in contrast to a pure Rashba or a pure Dresselhaus system the frequency range where $\mathrm{Re}\sigma_{ab}(\omega)\neq0$ is more broadened $\hbar\omega_-\leq\hbar\omega\leq\hbar\omega_+$, where

\begin{equation}\label{omegamax}
\hbar\omega_\pm=\hbar\Omega_\pm(\Delta_\pm)=2\Delta_\pm\left(\sqrt{m^2\Delta_\pm^2+2mE_F}\pm m\Delta_\pm\right).
\end{equation}

\noindent To summarize, the conductivity of a spin-orbit coupled two-dimensional electron gas is determined by the formula (\ref{cond5}) and is non-zero in a tiny region $\hbar\Omega_-(\theta)\leq\hbar\omega\leq\hbar\Omega_+(\theta)$. Interestingly, the integral in (\ref{cond5}) can be done in closed analytical form

\begin{equation}
\mathrm{Re}\sigma_{ab}(\omega)=\frac{e^2}{16\pi\hbar}\left(\begin{array}{cc}
I_3 & I_2 \\
I_2 & I_1
\end{array}\right).
\end{equation}

\noindent For the sake of brevity, we have introduced

\begin{gather}
I_1(\zeta)=\frac{h}{\sqrt{fh-g^2}}\arctan\left(\frac{g+h\tan\zeta}{\sqrt{fh-g^2}}\right)+\frac{g+h\tan\zeta}{f+2g\tan\zeta+h\tan^2\zeta},
\end{gather}

\begin{equation}
I_2(\zeta)=\frac{g}{\sqrt{fh-g^2}}\arctan\left(\frac{g+h\tan\zeta}{\sqrt{fh-g^2}}\right)+\frac{f+g\tan\zeta}{f+2g\tan\zeta+h\tan^2\zeta},
\end{equation}

\begin{equation}
I_3(\zeta)=\frac{f}{\sqrt{fh-g^2}}\arctan\left(\frac{g+h\tan\zeta}{\sqrt{fh-g^2}}\right)+\frac{fg+(2g^2-fh)\tan\zeta}{h\left(f+2g\tan\zeta+h\tan^2\zeta\right)},
\end{equation}

\noindent then, the components of the conductivity tensor (\ref{cond5}) are to be found from

\begin{equation}
I_i=\left\lbrace\begin{array}{cc}
I_i(\zeta^-_+)-I_i(\zeta^-_-), & \omega_-\leq\omega\leq\omega_a \\ \\
I_i(\zeta^-_+)-I_i(\zeta^+_+)+I_i(\zeta^+_-)-I_i(\zeta^-_-), & \omega_a\leq\omega\leq\omega_b \\ \\
I_i(\zeta^+_+)-I_i(\zeta^+_-), & \omega_b\leq\omega\leq\omega_+
\end{array}\right.
\end{equation}

\noindent where $\tan\zeta_\pm^{(q)}=\frac{-g\pm\sqrt{g^2-\left(f-d^{(q)}\right)\left(h-d^{(q)}\right)}}{h-d^{(q)}}$, $d^\pm=\frac{\hbar^2\omega^2}{4m\left(2E_F\pm\hbar\omega\right)}$, and $\hbar\omega_a=\hbar\Omega_+(\Delta_-)$, $\hbar\omega_b=\hbar\Omega_-(\Delta_+)$.

\end{document}